# A simple drain current model for Schottky-barrier carbon nanotube field effect transistors

D. Jiménez, X. Cartoixà, E. Miranda, J. Suñé, F. A. Chaves, and S. Roche

D. Jiménez, X. Cartoixà, E. Miranda, J. Suñé, and F. A. Chaves are with the Departament d'Enginyeria Electrònica, Escola Tècnica Superior d'Enginyeria, Universitat Autònoma de Barcelona, 08193-Bellaterra, Barcelona, Spain.

S. Roche is with DSM/DRFMC/SPSMS, Commisariat à l'Energie Atomique, France

Corresponding author: david.jimenez@uab.es

***Abstract-*** We report on a new computational model to efficiently simulate carbon nanotube-based field effect transistors (CNT-FET). In the model, a central region is formed by a semiconducting nanotube that acts as the conducting channel, surrounded by a thin oxide layer and a metal gate electrode. At both ends of the semiconducting channel, two semi-infinite metallic reservoirs act as source and drain contacts. The current-voltage characteristics are computed using the Landauer formalism, including the effect of the Schottky barrier physics. The main operational regimes of the CNT-FET are described, including thermionic and tunnel current components, capturing ambipolar conduction, multichannel ballistic transport and electrostatics dominated by the nanotube capacitance. The calculations are successfully compared to results given by more sophisticated methods based on non-equilibrium Green's function formalism (NEGF).

# 1. Introduction

In recent years, the interest in novel device structures able to surmount the miniaturization limits imposed by silicon-based transistors has led researchers to explore alternative technologies such as those originated in the field of carbon nanotubes. Carbon nanotubes are a very promising material for future nanoelectronics, both as interconnects and as critical elements for field-effect transistors because of their low dimensionality and resulting impressive electronic properties. For a given diameter, either metallic or semiconducting nanotubes are found depending on the topological arrangements of carbon atoms in the nanotube matrix [1]. By engineering semiconducting carbon nanotubes-based field effect transistors (CNT-FET) excellent device performances have been obtained [2,3]. Indeed, owing to low backscattering, ballistic transport is sustained up to micron scale (for reasonable values of bias voltages), which provides high-density on-state current capability, whereas the large and diameter-dependent energetic spacing between one-dimensional subbands results in tunable energy gaps, enabling the CNT-FET to operate with a very low off-state current. Concerning the injection regime, it has been shown that, due to an unconventional electrostatics and Fermi level pinning effect [4], the nanotube/metal contact feature can be tuned from ohmic to Schottky-like behavior. In particular, it was shown that palladium and rhodium metals establish ohmic contacts to semiconducting tubes with diameter larger than ~1.6nm, whereas non-negligible Schottky barrier heights were found with tubes with diameter smaller than ~1.6nm [5].

At present, an important issue is the development of accurate and efficient computational tools for simulations of carbon nanotube-based devices. These models should be ultimately designed to capture the essential physics at a moderate cost, while properly describing the observed phenomenology in CNT-FETs under different biasing conditions. The aim of these models is to serve as guidelines for understanding the ongoing experimental work at this early stage of development, and for first-order design and projection purposes. Several simple approaches have been reported for simulating CNT-FETs assuming ohmic contacts between the semiconducting carbon nanotube and drain/source electrodes [6,7], in analogy with the conventional silicon transistor. For these devices, the potential barrier between the electrodes and the channel is modulated by the gate electrode allowing thermionic current to pass (on-state) or not (off-state). An alternative device concept is the Schottky barrier CNT-FET. This is the most common design because the ion implantation techniques employed to create a doping profile as in conventional semiconductors cannot be used for nanotubes. The reason being that removing any carbon atoms forming the nanotube and replacing them by a dopant seriously affects the nanotube properties [8]. For this device concept, the Schottky barrier width is modulated by the gate voltage, changing the tunnelling probability for carriers to flow along the channel. In this work, we focus on the electrical modeling of (single-wall) carbon nanotube field effect transistor (CNT-FET) with Schottky barrier injectors (Fig. 1). Specifically, we have developed a physics-based analytical model for the current-voltage (I-V) characteristics that captures the main features exhibited by these structures, such as: (a) thermionic and tunnel emission [9]; (b) ambipolar conduction [10]; (c) ballistic transport [11]; (d) multimode propagation [12]; and (d) electrostatics dominated by the nanotube capacitance [13,14]. The main point of our approach is the computational implementation simplicity in comparison with more rigourous treatments such as the Non-Equilibrium Green's Functions method [15].

## 2. The model

The CNT-FET is formed by three distinctive regions which reflect the variation of the energy band diagram along the device: two injecting regions at the ends of the nanotube and a central region where ballistic transport occurs (Fig. 2). If the electrodes are sufficiently separated (long-channel hypothesis), this latter region is flat band type, its energetic level being essentially determined by the nanotube capacitance, which is assumed to dominate over the insulator capacitance. This condition is often referred to as the quantum capacitance limit [14]. Under this condition, the nanotube capacitance and associated storaged charge are close to zero. Accordingly, the gate capacitance is dominated by the nanotube capacitance, yielding a surface potential $\varphi_S=qV_{GS}$, where $V_{GS}$ is the gate-source applied voltage and $q$ the electron charge. The spatial band diagram in the vicinity of the metal/nanotube interfaces can then be analytically solved using the 1D Laplace equation along the transport direction (see Appendix). For the coaxial gate geometry, the bottom of the conduction band profile at the left ($E_L$) and right ($E_R$) regions close to the metal reservoirs can be written respectively as:

$$E_L(z) = \varphi_{SB} - q(V_{GS} - V_{FB})\left(1 - e^{\frac{-2z}{t_{ox}}}\right) \quad (1)$$

$$E_R(z) = (\varphi_{SB} - qV_{DS}) - q(V_{GS} - V_{FB})\left(1 - e^{\frac{2(z-L)}{t_{ox}}}\right) \quad (2)$$

where $\varphi_{SB}$ is the Schottky barrier height at the metal-nanotube contact, $V_{FB}$ the flat-band voltage, $t_{ox}$ the insulator thickness and L the total length of the nanotube. Note that, due to the chosen energy reference system, the top of the Schottky barrier at the drain contact depends on the drain bias (=$\varphi_{SB}$-$qV_{DS}$). The valence band has the same profile as the conduction band but downshifted by a value equal to the nanotube energy gap $E_g$.

The current through the structure is calculated by means of the Landauer-Büttiker formula assuming a one-dimensional ballistic channel in between contacts that are further connected to external reservoirs, where dissipation takes place:

$$I = \frac{4q}{h}\sum_n \int_{-\infty}^{\infty} \text{sgn}(E)T_n(E)\big(f(\text{sgn}(E)(E-E_{FS})) - f(\text{sgn}(E)(E-E_{FD}))\big)dE \quad (3)$$

where *h* is the Planck's constant, f(E) the Fermi-Dirac distribution function, $T_n$ the transmission probability of the $n^{th}$ subband (*n* starting from zero), and *sgn* the sign function (introduced for expressing both electron and hole currents in a compact manner). This expression accounts for the spin degeneracy of the injected carriers. As it is well known, within this formalism, the transmission probability plays a central role. In order to compute the total current, the current of each subband is split into two components, namely the thermionic and tunneling currents, and then added as indicated by (3). In the case of the thermionic current (electrons injected with energies E>max($\varphi_{SB}$+|$\Delta E_n$|,$\varphi_{SB}$-q$V_{DS}$+|$\Delta E_n$|) and holes injected with energies E<min($\varphi_{SB}$-$E_g$+|$\Delta E_n$|,$\varphi_{SB}$-q$V_{DS}$-$E_g$+|$\Delta E_n$|) we assume $T_n(E)=1$, where $\Delta E_n$ is the difference between the energies of $n^{th}$ subband ($E_n$) and the lowest subband (see Eq. (9)). For the tunneling component, the transmission through a single Schottky barrier is computed using the Wentzel-Kramers-Brillouin (WKB) approximation:

$$T_n(E) = \exp\left(-2\int_{z_i}^{z_f} k_z(z)dz\right) \quad (4)$$

where $z_i$ and $z_f$ are the classical turning points. For example, for electrons injected from the source to the nanotube conduction band, the classical turning points are determined by the following relations:

$$z_i = 0;$$
$$\varphi_{SB} - E_L(z_f) = E, \quad if \quad \varphi_{SB} \geq E > \varphi_{SB} - E_g;$$

$$\varphi_{SB} - E_L(z_i) - E_g = E;$$
$$\varphi_{SB} - E_L(z_f) = E, \quad if \quad E \leq \varphi_{SB} - E_g, \quad (5)$$

where $E_L(z)$ is given from (1). The parallel momentum $k_z(z)$ is linked to the energy through the *E-k* dispersion relationship of the carbon nanotube, which can be expressed as [14,16]:

$$E - E(z) = \pm \frac{3d}{2} V_{pp\pi} \sqrt{k_n^2 + k_z^2} \quad (6),$$

where E(z) is the energy profile of the conduction (valence) band for electrons (holes), $k_n$ is the component of the momentum perpendicular to the transport direction for the $n^{th}$ subband; $V_{pp\pi}$=2.97 eV is the nearest-neighbor interaction parameter [16] and d=0.142 nm is the carbon-carbon bond distance [17]. The sign "+" and "-" of the RHS of (6) applies to the conduction and valence band respectively. For a nanotube with a chiral vector given by ($n_1$,$n_2$), the perpendicular momentum $k_n$ is given by [16]:

$$k_n = \frac{|3n - n_1 + n_2|}{3R} \quad (7),$$

where R is the carbon nanotube radius:

$$R = \frac{d}{2\pi} \sqrt{3(n_1^2 + n_2^2 + n_1 n_2)} \quad (8),$$

and the minimum of the $n^{th}$ subband can be written as:

$$|E_n| = \frac{|3n - n_1 + n_2|}{2R} V_{pp\pi} d \quad (9).$$

The energy gap of a semiconducting carbon nanotube ($n_1$-$n_2$≠3n) can be easily determined from (9) resulting in $E_g$=$V_{pp\pi}$d/R. From (4) and (6), the transmission coefficient through a single Schottky barrier can be expressed as:

$$T_n(E) = \exp\left(-\int_{z_i}^{z_f} \frac{2E_g}{3dV_{pp\pi}} \sqrt{\left(\frac{3dV_{pp\pi} k_n}{E_g}\right)^2 - \left(\frac{E - E(z)}{E_g/2}\right)^2} dz\right) \quad (10)$$

Eq. (10) must be numerically evaluated for each energy. Since the CNT-FET is an ambipolar device, in general, there are four possible current contributions: electron/hole thermionic currents and electron/hole tunneling currents (Fig. 3). In addition, depending on the magnitude of the applied bias, multiple reflections can arise between the series combination of both Schottky barriers, in the same way as it occurs in a Fabry-Pérot cavity [18]. Fig. 4 illustrates this phenomenon for the case of half-gap Schottky barriers contacts. If the bias point satisfies $V_{GS}>V_{DS}$, a double tunneling barrier for electrons is formed. In this case, neglecting phase coherence, the transmission coefficient is given by the following expression [19]:

$$T_n(E) = \frac{T_{Ln}T_{Rn}}{T_{Ln}+T_{Rn}-T_{Ln}T_{Rn}} \qquad (11),$$

where $T_{Ln}(E)$ and $T_{Rn}(E)$ are the transmission coefficients for the left and right barriers associated with the $n^{th}$ subband, respectively. We have checked, by solving numerically the 1D Schrödinger equation with the analytical energy profile given by (1) and (2), that neglecting the phase coherence does not appreciably affect the I-V characteristics in the typical range of temperatures where transistors operate. Even though the spectrum of the transmission is averaged when phase coherence is disregarded leading to a smoothed current density spectrum, the same current is obtained if the tail of Fermi-Dirac distribution is large enough. Only at low temperature conditions, out of the range from the conventional operational window, phase coherence would need to be incorporated in the model.

## 3. Model assessment

In this Section, our simulations are compared with those provided by self-consistent quantum mechanical simulations based on the non-Equilibrium Green's Functions (NEGF) method [15]. In particular, we have tested a nominal device consisting of a long channel CNT-FET with a (13,0) semiconductor carbon nanotube and insulator thickness of 2 nm, with a half-gap Schottky barrier height. More thoroughly, the test compares the transfer characteristics, showing the effect of power supply voltage scaling, nanotube diameter scaling, and barrier height dependence. The transfer characteristics exhibit two branches at the left and right of the minimum point; the so-called ambipolar conduction point. This minimum occurs at $V_{GS}=V_{DS}/2$ for a half-gap Schottky barrier height ($\varphi_{SB}=E_g/2$). In this case, the spatial band diagram is symmetric for electrons and holes and the electron and hole tunneling currents are identical. When $V_{GS}$ is greater (smaller) than $V_{DS}/2$, the Schottky barrier width for electrons (holes) is reduced, and the electron (hole) tunneling current dominates. The effect of power supply up-scaling is to further reduce the Schottky barrier width making it more transparent and allowing more turn-on current to pass through (Fig. 5). Besides, downsizing the nanotube diameter increases the gap between the conduction and valence bands. Hence, the current is smaller because high-energy levels in the conduction and valence bands become less populated, in accordance with the Fermi-Dirac distribution (Fig. 6). On the other hand, the resulting on-off current ratio, a figure-of-merit for digital circuits, is largely improved. The deviation of the Schottky barrier heigth respect to the half-gap case results in a shift of the ambipolar conduction point and asymmetries between the left and right branches of the transfer characteristic (Fig. 7). Notice the close agreement between the proposed simple model and NEGF's simulator. It is worth pointing out that the proposed model remains valid as long as the CNT-FET operates at the quantum capacitance limit. To get further insight into this issue, we have simulated a CNT-FET with a thicker insulator ($t_{ox}=40$ nm). Now, the oxide capacitance is comparable with the nanotube capacitance, and the central region of the band diagram does not rigidly move with $V_{GS}$. The energy level of the bottom of the conduction band should be determined by the capacitive divider formed by the series combination of the insulator and nanotube capacitances. For example, a CNT-FET with a half-gap Schottky barrier height operating at the quantum capacitance limit largely overestimates the current, as can be clearly seen in Fig. 8. Finally, we report on the output characteristics of the Schottky barrier CNT-FET (Fig. 9). Two different conduction mechanisms at the left and right of the ambipolar conduction point $V_{DS}=2V_{GS}$ are observed. For $V_{DS}<2V_{GS}$, the conduction mechanism is electron tunneling. Specifically, when $V_{DS}<V_{GS}$, Schottky barriers (for electrons) close to the source and drain electrodes arise and multiple reflections take place between the two barriers. When $V_{DS}$ is slightly greater than $V_{GS}$, the Schottky barrier (for electrons) close to the drain contact fades out, and the Schottky barrier (for holes) close to the drain contact automatically develops. Then, a current saturation is observed. The reason being that $V_{DS}$ does not have control over the Schottky barrier width close to the source contact (provided that short-channel effects are negligible, in accordance with the long-channel hypothesis). At the right part of the ambipolar conduction point ($V_{DS}>2V_{GS}$), the Schottky barrier width for holes close to the drain contact becomes very thin and hole tunneling current dominates.

## 4. Conclusions

In conclusion, we have developed a simple model for the I-V characteristics of Schottky-barrier carbon nanotube field effect transistors which captures the main physical effects governing the operation of this device. The results obtained applying this model to prototype devices are in close agreement with a more rigorous treatment based on the NEGF approach, thus validating the approximations made. The present model can assist at the design stage and when projecting the ultimate performance limits, as well as for quantitative understanding of the experimental work.

## Acknowlegdments

This work was supported by the Ministerio de Ciencia y Tecnología under project TIC2003-08213-C02-01 and the European Commission under Contract 506653. ("EUROSOI").

**Appendix**

Let us consider a coaxial gate geometry, with the carbon nanotube a one-dimensional object (radius zero), located at the axis of the structure (r=0). In the quantum capacitance limit, the charge in the nanotube is zero and the electrostatics of this system is governed by Laplace's equation:

$$\frac{1}{r}\frac{\partial}{\partial r}\left(r\frac{\partial E(r,z)}{\partial r}\right)+\frac{\partial^2 E(r,z)}{\partial z^2}=0 \quad (A1),$$

where E(r,z) is the potential energy. We will assume that in the radial direction the potential energy is parabolic type:

$$E(r,z)=a(z)r^2+b(z) \quad (A2),$$

with a(z) and b(z) z-dependent functions. The linear term in *r* was not included because of symmetry considerations. The boundary condition at the surface of the cylindrical structure is $E(r=t_{ox},z)=\varphi_{SB}-(V_{GS}-V_{FB})$. Replacing (A2) into (A1), we can write the following effective 1D Laplace's equation at the cylinder axis (r=0):

$$\frac{d^2 a(z)}{dz^2}-\frac{a(z)}{\lambda^2}=0 \quad (A3).$$

with $\lambda=t_{ox}/2$ a characteristic length that can be interpreted as a measure of the Schottky barrier width. Eq. (A3) has the general solution $a(z)=A\exp(-z/\lambda)+B\exp(z/\lambda)$. For determining A and B, the following two boundary conditions must be satisfied:

$$E(r=0,z=0)=\varphi_{SB}$$
$$E(r=0,z=\infty)=\varphi_{SB}-(V_{GS}-V_{FB}) \quad (A4)$$

yielding the potential energy profile along the center of the system:

$$E(r=0,z)=E(z)=\varphi_{SB}-(V_{GS}-V_{FB})(1-\exp(-z/\lambda)) \quad (A5).$$

**Figure captions**

**Figure 1.** CNT-FET cross-section. The electrodes are assumed to be metallic and the body is a semiconductor carbon nanotube.

**Figure 2.** Spatial band diagram along the transport direction.

**Figure 3.** Spatial band diagram at the ambipolar conduction point showing the four components forming the current.

**Figure 4.** Spatial band diagram for a half-gap Schottky barrier CNT-FET polarized at $V_{GS}>V_{DS}$, showing two reflecting barriers for the electrons.

**Figure 5.** Effect of the power supply voltage scaling on the transfer characteristics for the nominal device.

**Figure 6.** Effect of the nanotube diameter scaling on the transfer characteristics. The thermionic current for electrons (holes) for the (25,0) carbon nanotube is shown. Note that it is about 3-4 orders of magnitude smaller than the tunnelling current.

**Figure 7.** Effect of the barrier height on the transfer characteristics for the nominal device.

**Figure 8.** Transfer characteristics for a (25,0) CNT-FET with a thick insulator thickness ($t_{ox}$=40 nm), illustrating the operation far from the quantum capacitance limit.

**Figure 9.** Output characteristics for the nominal device.

**Figure 1**

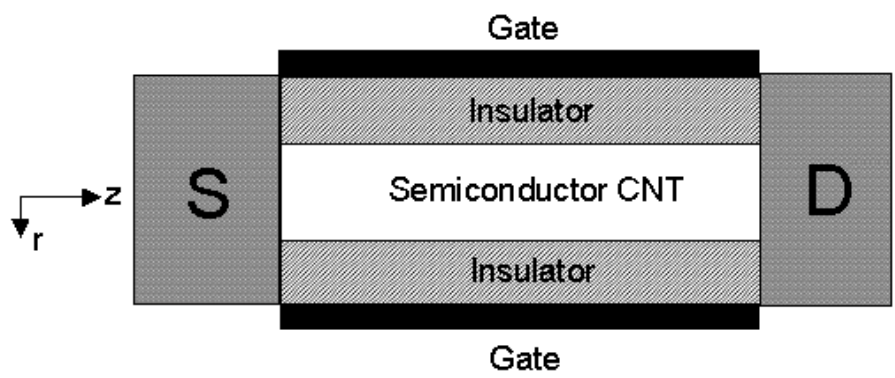

**Figure 2**

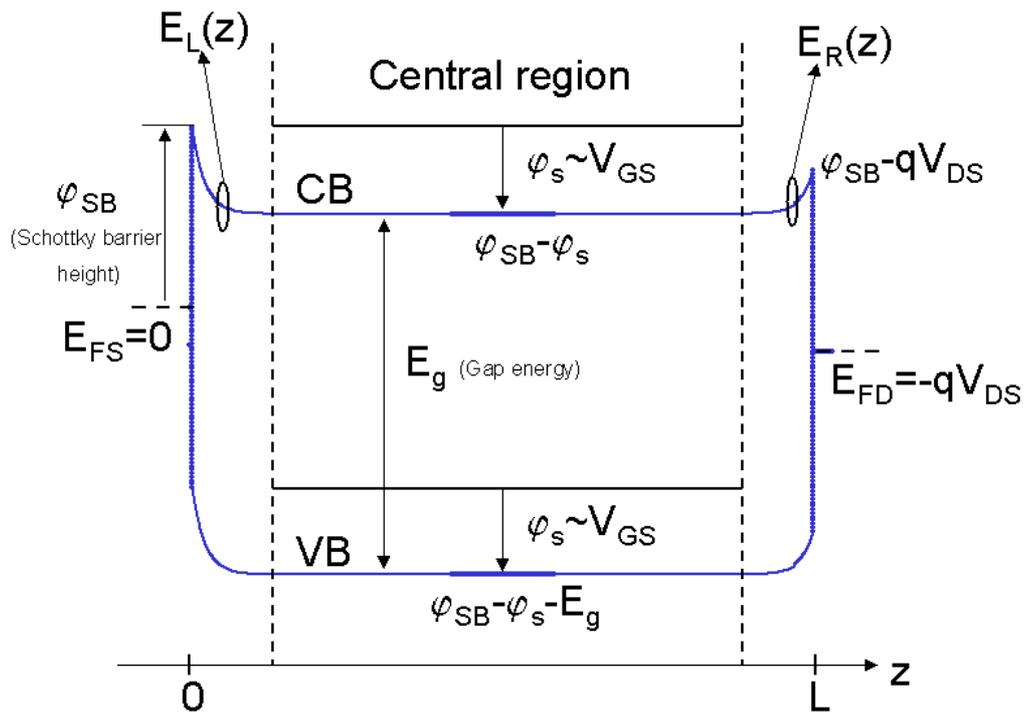

**Figure 3**

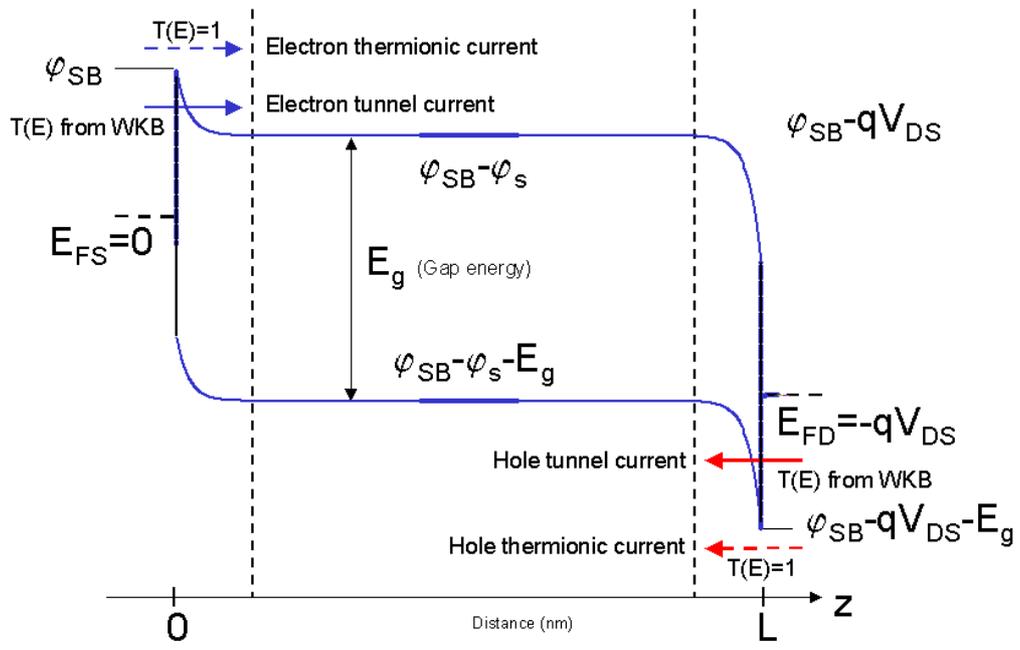

**Figure 4**

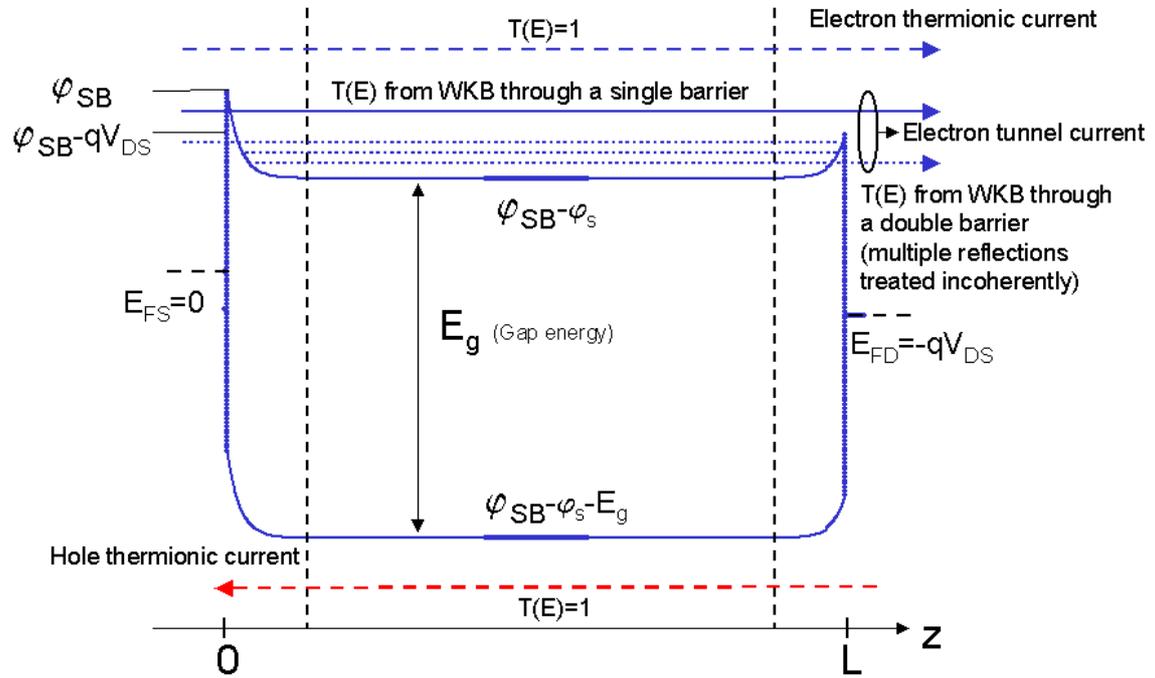

**Figure 5**

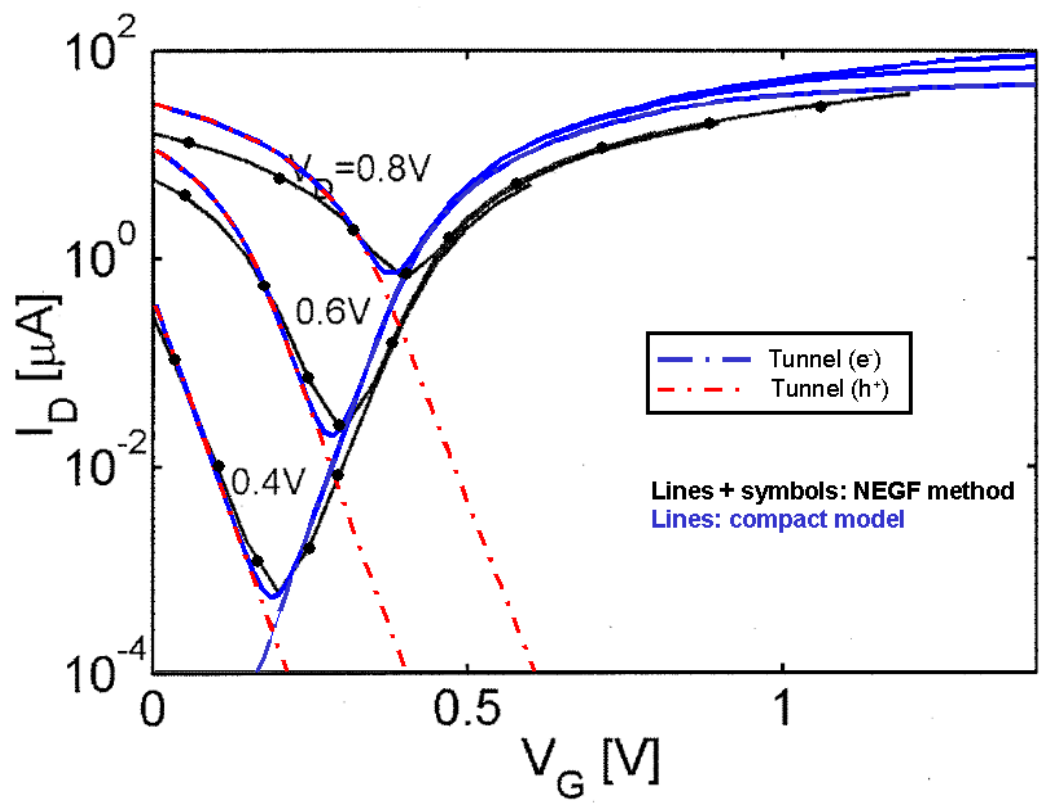

**Figure 6**

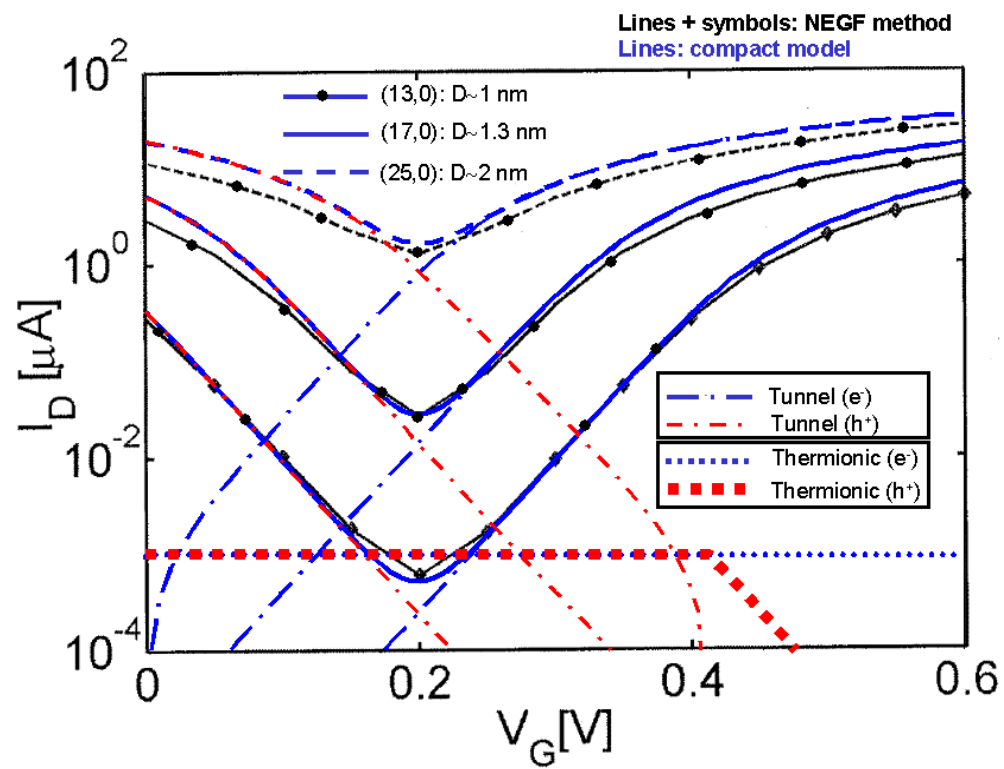

**Figure 7**

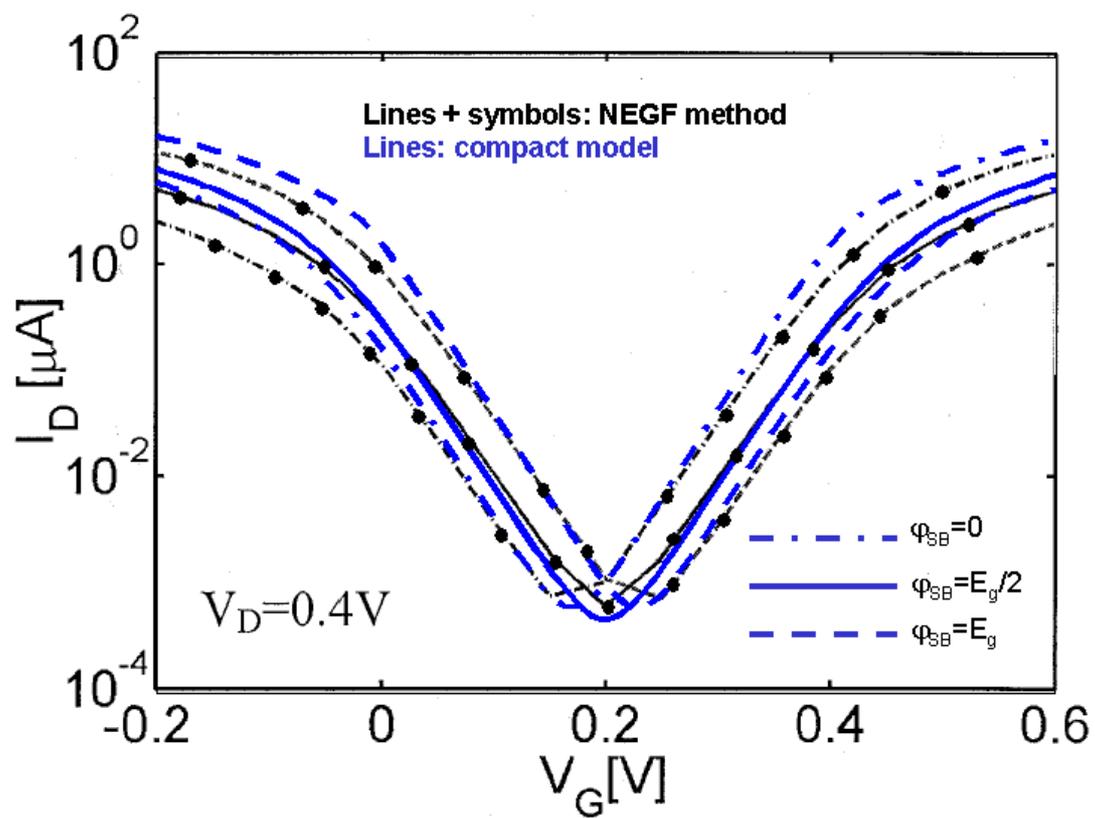

**Figure 8**

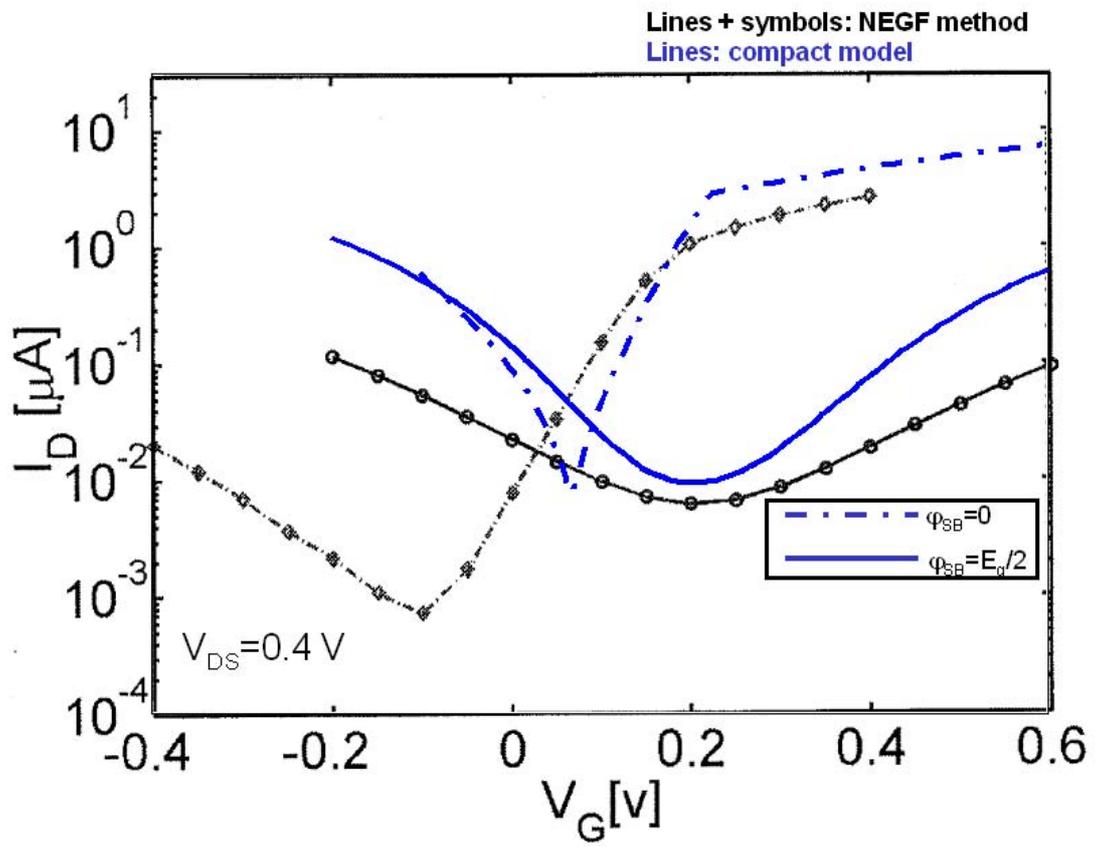

**Figure 9**

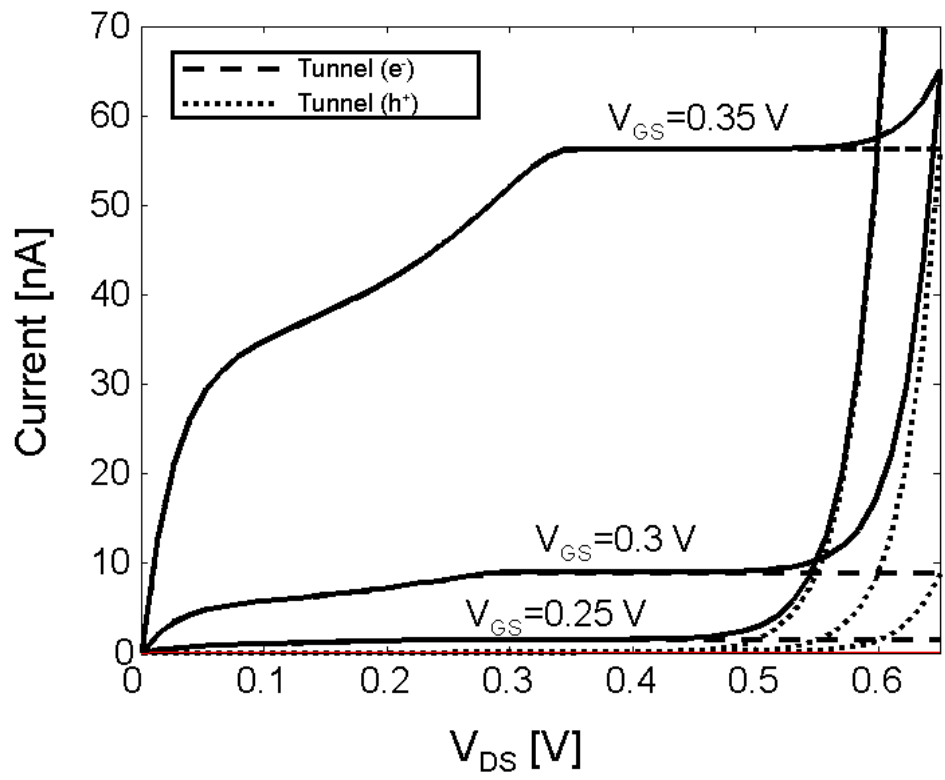